\documentclass[12pt]{iopart}
\usepackage{graphicx}
\usepackage{amssymb}

\graphicspath{{EPS/}}
\usepackage{color}

\newcommand{\begeq}{\begin{equation}}           
\newcommand{\eeq}{\end{equation}}
\newcommand{\begeqn}{\begin{eqnarray}}
\newcommand{\eeqn}{\end{eqnarray}}

\newcommand{\up}{\uparrow}
\newcommand{\down}{\downarrow}
\newcommand{\G}{\Gamma}
\newcommand{\s}{\sigma}

\newcommand{\D}{\Delta}
\newcommand{\w}{\omega}

\newcommand{\summ}{\sum\nolimits}
\newcommand{\epsd}{\varepsilon_d}

\begin{document}

\title[NRG calculation of the spectral function of the SC-AM]{Numerical renormalization group calculation of near-gap peaks in spectral functions of the Anderson model with superconducting leads}

\author{T Hecht$^{1}$, A Weichselbaum$^{1}$, J von Delft$^{1}$ and R Bulla$^{2,3}$}
\address{$^{1}$ Physics Department, Arnold Sommerfeld Center for Theoretical Physics and Center for NanoScience, Ludwig-Maximilians-Universit\"{a}t M\"{u}nchen, Germany}
\address{$^{2}$ Theoretische Physik III, Elektronische Korrelationen und Magnetismus, Universit\"at Augsburg, Germany}
\address{$^{3}$  Institut f\"ur Theoretische Physik, Universit\"at zu K\"oln, Germany}

\ead{Theresa.Hecht@physik.uni-muenchen.de}

\begin{abstract}
We use the numerical renormalization group method (NRG) to
investigate a single-impurity Anderson model with a
coupling of the impurity to a superconducting host.
Analysis of the energy flow shows,
in contrast to previous belief, that NRG iterations
can be performed up to a large number of sites,
corresponding to energy differences far below the
superconducting gap $\Delta$. This allows us to
calculate the impurity spectral function $A(\omega)$
very accurately for frequencies
$\vert\omega\vert\sim\Delta$, and to resolve,
in a certain parameter regime, sharp peaks in $A(\omega)$
close to the gap edge.
\end{abstract}
\pacs{75.20.Hr, 74.50.+r}



\section{Introduction}

The vast progress in nanofabrication during the last decades made it possible to study basic physical effects in a very controlled manner.
One example of such highly controllable devices are quantum dots  \cite{Kouwenhoven1998}, 
which are used, amongst various applications, for a detailed and very controlled study of the Kondo effect  \cite{Kondo1964,Glazman1988,Ng1988,Goldhaber-Gordon1998}, which is one of the prime examples of many-body phenomena.
Below a critical temperature (Kondo temperature $T_K$) a local moment, provided by the spin of an electron occupying the quantum dot, gets screened by reservoir electrons within an energy window $T_K$ around the Fermi energy.

These seminal experimental works on the Kondo effect, together with the possibility to engineer reservoir properties, raise the following intriguing question: What interesting effects may arise if the local moment in the quantum dot is coupled to superconducting leads while parameters like the Kondo temperature or the superconducting gap can be adjusted arbitrarily? 
In a  superconductor as described by Bardeen, Cooper and Schrieffer (BCS) \cite{Bardeen1957}, electrons with opposite spin and momentum form Cooper pairs, thereby expelling  the lead density of states  around the Fermi energy. An energy gap occurs.
Obviously, when combining a Kondo quantum dot with superconducting reservoirs both effects compete: Screening of the local moment by electrons around the Fermi energy against pair formation of the latter.

This competition  has attracted a lot of interest and various methods were used to analyze the  properties of the system in the above-mentioned limits, see for example the references in \cite{Balatsky2006} or \cite{Glazman1989,Shiba1973,Clerk2000,Avishai2003}. 
In the early 1990's  Satori \emph{et al.}\ \cite{Satori1992} proposed to apply the numerical renormalization group method (NRG) \cite{Wilson1975, Bulla2008} to the problem.
Since then NRG was used to calculate the ground state and subgap bound state properties like position and degeneracy \cite{Satori1992,Yoshioka2000,Tanaka2007} as well as their spectral weight \cite{lim-2006,Bauer2007}.
Also dynamic quantities like the spectral function \cite{Bauer2007} or the Josephson current \cite{Oguri2004,Karrasch2007} can be calculated with NRG, as was done recently.
 
The first main goal  of the present paper is to gain insight into the way  NRG works when applied to a system of one local level coupled to a superconducting lead. 
NRG is a well established method for solving strongly correlated impurity problems.
 Actually, it was invented to solve the Kondo problem \cite{Wilson1975} and since then has been generalized to various schemes  involving localized states coupled to fermionic \cite{Krishnamurthy1980} or bosonic \cite{Bulla2003,Bulla2005} baths. For a review, see \cite{Bulla2008}.
The key idea of NRG is to discretize the conduction band logarithmically. This leads to a chain Hamiltonian with exponentially decreasing couplings, the so called Wilson chain. It can be solved iteratively by enlarging the system site by site: Due to the decreasing couplings every new site can be treated as a perturbation of the old system, thus increasing the resolution with every step.

It is, however, not at all obvious whether NRG still works for superconducting leads at energy resolutions well below the gap. This is because the pairing energy $\Delta$ is the same at all energy scales, thus remains a constant on-site contribution also in the above-mentioned chain structure.
At iterations where the coupling of the Wilson chain reaches $\Delta$, it is not obvious whether  added sites still can be understood  as a perturbation in the iterative process  or not,  that is whether NRG still works or not.

In order to address this problem, we analyze in detail the flow of the eigenenergies during the NRG procedure. We show that  NRG is indeed capable to resolve the continuum close to the gap without any restriction on the energy scale of the superconducting gap.

The second goal of this work consists in the calculation of the impurity spectral function close to the gap edge at zero temperature.
As in the Anderson model (with normal leads) a Kondo resonance may form. 
However, due to the superconducting property of the  leads a gap opens up around the Fermi energy, cutting the resonance. 
Our main interest lies in the study of the continuum contribution to the spectral function close to the gap edge, implying the need of high resolution in that regime. 
Our calculations cover not only the regime $\Delta\gtrsim T_K$, for which the continuum part of the spectral function was studied in \cite{Bauer2007}, but also $\Delta\ll T_K$.
In the latter regime we find a sharp peak at the gap edge, vastly exceeding the Kondo resonance contribution.
We expect this to lead to an enhanced linear conductance, as  observed in a recent experiment \cite{Buitelaar2002}
with carbon nanotube quantum dots coupled to superconducting leads.
They report dramatic enhancement of the linear conductance (only)  in the regime $\Delta\ll T_K$.

We find similar behaviour of the spectral function in the non-interacting case where an analytical solution exists. We analyze this solution and compare our NRG results against it. We find excellent agreement especially at energies close to the gap, as expected from our study of the energy spectrum.
\\

The paper is organized as follows:
In section 2  we introduce the superconducting-lead Anderson model (SC-AM) and the NRG method.  In section 3 the flow of the energy spectrum of the SC-AM is analyzed. NRG is shown to work at resolutions well below the energy scale of the superconducting gap. 
Section 4  discusses the calculation of the spectral function. In the last part we conclude and summarize our results.


\section{The model and the method}
In this section we introduce our model for the quantum dot coupled to a left and right superconducting  reservoir, as well as the method we use, the numerical renormalization group method  \cite{Wilson1975,Krishnamurthy1980}. We perform a Bogoliubov as well as a particle-hole transformation. We derive the formulas in a general form. For convenience, the discussions in the later sections will be restricted to a real order parameter of the superconductors, equivalent to only one lead.
In the non-interacting limit the system can be understood in terms of a simple single-particle  picture. The latter can be solved exactly and will serve as a guideline for gaining a deeper understanding of the problem.
%

%

\subsection{Superconducting-lead Anderson model (SC-AM)}
To describe a quantum state coupled to two  superconducting reservoirs, 
we consider the standard Anderson model (AM) for a local level coupled to two metallic, non-interacting reservoirs \cite{Anderson1961}, and add a BCS-type term 
 $H_\D$,  describing pair formation in the leads. This SC-lead Anderson model, to be called SC-AM, is then described by
\begeq
   H=H_{dot}+H_{hyb}+H_{lead}+H_{\D},
   \label{H}
\eeq
with
\begeqn
   H_{dot}
   &=& \sum_\sigma \varepsilon_d n_{d\sigma} + 
   U n_{d\up} n_{d\down} 
   \label{H_d}
   \\
   H_{hyb}
   &=& \sum_{l=L,R}\sum_{\bf k \s}V_l(c^\dag_{l\bf k \s}d_\s + d_\s^\dag c_{l\bf k\s})
   \label{H_hyb}
   \\
   H_{lead}
   &=& \sum_{l=L,R}\sum_{\bf k \s}\varepsilon_{\bf k} c^\dag_{l \bf k \s}c_{l\bf k \s}
   \label{H_l}
   \\
   H_{\D}
   &=& -\sum_{l=L,R}\sum_{\bf k} 
      \Delta_l( e^{i\phi_l}c^\dag_{l\bf k \up}c^\dag_{l\bf -k \down}
               +e^{-i\phi_l}c_{l\bf- k \down}c_{l\bf k \up}).
   \label{H_BCS}
\eeqn
Electrons on the dot with spin $\sigma=\{\up,\down\}$ are created by $d^\dag_\sigma$ and interact via the Coulomb repulsion $U$ with each other, $n_{d\s}=d^\dag_\s d_\s$ being the charge operator for spin $\s$.
In the hybridization term $H_{hyb}$, the coupling strength $V_l$ between dot and lead states is assumed to be real and independent of the wave vector $\bf k$. 
$c^\dag_{l\bf k\s}$ creates an electron in lead $l=L,R$, respectively.
$H_{lead}$  describes the conduction band of metallic leads. 
We assume an isotropic and linearized dispersion. 
The density of states is then constant, $\rho_0=1/2D$,  where the band ranges from $-D$ to $D$. In the following $D=1$ will serve as energy unit.
The pairing of electrons with opposite spin and momentum (Cooper-pairs) is described by $H_\D$.
 $\Delta_l$ is the magnitude (later on often called the gap), $\phi_l$  the phase of the order parameter of the superconductor.
For simplicity, we assume left-right symmetry, i.e.\  $\Delta_l=\Delta$, $\phi_l=\pm \phi/2$ and $V_l=V$, where $l=L,R$, respectively. Then the unitary rotation %
\begeqn
  \left(
     \begin{array}{l}
       c_{e\bf k\s}\\
       c_{o\bf k\s}
     \end{array}
   \right)
   =
   \,
   \frac{1}{\sqrt2}
   \left(
     \begin{array}{lr}
       e^{-i\phi/4}&e^{i\phi/4}\\
       -i e^{-i \phi/4} & i e^{-i\phi/4}
     \end{array}
   \right)
   \,
   \left(
     \begin{array}{l}
       c_{L\bf k\s}\\
       c_{R\bf k\s}
     \end{array}
   \right)\,\,\,
   \nonumber
\eeqn
of the reservoir operators yields a real Hamiltonian.
%

It is well known \cite{Bogoliubov1958,Valatin1958}  that the Hamiltonian of a bulk superconductor [equations (\ref{H_l}) and (\ref{H_BCS})]
can be diagonalized by a Bogoliubov transformation.
Note, though, that the SC-AM cannot be understood as an AM  with a superconducting lead density of states ($\rho_\D=|\varepsilon_k|/\sqrt{\varepsilon_k^2+\D^2}$).
This is because the Bogoliubov transformation explicitly depends on $\bf k$. 
The hybridization term $c^\dag_{l\bf k\s} d_{\s}+d^\dag_\s c_{l\bf k \s}$ would transform to a complicated object that cannot be simplified by rotating the $d$ operators of the local dot space.

\subsection{The numerical renormalization group method (NRG)}
In the  1970's, K.G.\ Wilson came up with a  scheme for solving the Kondo problem nonperturbatively: the numerical renormalization group (NRG) \cite{Wilson1975}. 
Since then it was  generalized to various schemes, describing localized electronic states coupled to fermionic  \cite{Krishnamurthy1980} or bosonic \cite{Bulla2003,Bulla2005} baths. 
The NRG allows thermodynamic and dynamic properties of such strongly correlated systems to be calculated at zero as well as at finite temperature.
We first discuss the method for the AM ($\D=0$), then for $\D\neq 0$. For brevity  we apply all NRG transformations to the full SC-AM already in the discussion of the AM.

The key idea of NRG is to discretize the conduction band of the reservoir logarithmically.
The Hamiltonian can then be transformed to a chain Hamiltonian.
In this representation, equations (\ref{H_hyb}) to (\ref{H_BCS}) are mapped onto
\begeqn
   H_{hyb}
   &=& \sqrt{\frac{2\Gamma }{\pi}}\sum_{\sigma}
   \left[\cos{\frac{\phi}{4}}(f^\dag_{e0\s}d_\s + h.c.)
     -\sin{\frac{\phi}{4}}(f^\dag_{o0\s}d_\s + h.c.)\right]
   \label{H_hyb_NRG}
   \\
   H_{lead}
   &=& \frac{1}{2} \left(1+{\Lambda}^{-1} \right)
   \sum_{l=e,o}\sum_{\s}\sum_{n=0}^\infty \Lambda^{-n/2}\,\xi_n \,
   (f^\dag_{l n\s}f_{ln+1\s}+f^\dag_{l n+1\s}f_{ln\s})
   \label{H_l_NRG}
   \\
   H_{\D}
   &=& -\Delta \sum_n \left[(f^\dag_{en\up}f^\dag_{en\down}+h.c.)
     -(f^\dag_{on\up}f^\dag_{on\down}+h.c.)\right].
   \label{H_BCS_NRG}
\eeqn
Electrons on site $n$ in lead $l=e,o$ are created by $f^\dag_{nl\s}$.
The dot level only couples to the zeroth site of the so-called Wilson chain $H_{lead}$, where the hybridization is given by $\Gamma=\pi\rho 2V^2$ (the factor $2$ stems from the two leads).
$\Lambda>1$ is the discretization parameter of the conduction band. 
$\xi_n = (1-\Lambda^{-n-1})(1-\Lambda^{-2n-1})^{-1/2}(1-\Lambda^{-2n-3})^{-1/2} \approx 1$ for large $n$.
The hopping matrix elements between successive sites of $H_{lead}$ fall off exponentially with $\Lambda^{-n/2}$.  
The resulting energy scale separation  ensures that the AM can be solved iteratively. The recursion relation reads
\begeqn
   H_0
   &=&
   1/\sqrt{\Lambda} \left[ H_{dot}+H_{hyb}  
   -
   \Delta \sum_{l=e,o}\sum_\s
   s_l \left(f^\dag_{l0\up}f^\dag_{l0\down}+h.c.\right)\right]
   \nonumber
   \\
   H_{N+1}
   &=&
   \sqrt{\Lambda}\ H_N 
   +
   \frac{1}{2} \left(1+{\Lambda}^{-1} \right)\,
   \sum_{l=e,o}\sum_\s
   \xi_N 
   \left(f^\dag_{l N\s}f_{lN+1\s}+h.c.\right)
   \label{BCS:H_N+1}
   \\
   &&-
   \Delta \, \Lambda^{N/2}\sum_{l=e,o}\sum_\s
   s_l\left(f^\dag_{lN+1\up}f^\dag_{lN+1\down}+h.c.\right),
   \nonumber
\eeqn
ere $s_l=\pm 1$ for $l=e,o$, respectively.
The initial Hamiltonian of the system is related to the NRG Hamiltonian by
$
    H=\lim_{N\rightarrow \infty}
    \Lambda^{-(N-1)/2}\; 
    {H}_N.
$
This relation is  exact in the limit $\Lambda\rightarrow1$ and $N\rightarrow\infty$.

Sites are added successively  and at each step the enlarged system is diagonalized.
Each added site then acts as a perturbation of order $\Lambda^{-1/2}$ on the previous part of the chain.
Consequently, the typical energy resolution $\delta_n$ of the AM at iteration $n$ is given by $\delta_n^{{\rm AM}}\propto \Lambda^{-n/2}$.
Thus, by choosing the length $N$ of the chain large enough (so  that  $\Lambda^{-N/2}$ is much smaller than all other energies in the problem), all relevant energy scales can be resolved and treated properly. 
When adding a site to the system, the dimension of the Hilbert space gets multiplied by the dimension $d$ of the state space of that site, yielding
$d=4$  
for a single fermionic lead (empty, singly occupied (either up or down), doubly occupied).
Therefore  the dimension of the Hilbert space increases exponentially with the length of the chain. Wilson proposed a truncation scheme according to which only the lowest $N_{\rm kept}$ eigenstates are kept at each iteration, thereby  ensuring that the dimension of the truncated Hilbert space stays manageable.
Recently, it was shown that by  keeping track of the discarded states a complete, but approximate, basis of states can be constructed \cite{Anders2005,Anders2006}. 
This can be used to calculate dynamic properties like the spectral function $\mathcal A$ (see equation (\ref{BCS:A}) below) which rigorously satisfy relevant sum rules \cite{Peters2006,Weichselbaum2007}, like $\int d\omega \mathcal A(\omega) = 1$.

Applying the NRG mapping also to the pairing term of the SC-AM (as already done above), an on-site contribution appears, see equation (\ref{H_BCS_NRG}), 
constant in magnitude for each site. 
In the  limit $\Lambda^{-n/2}\gg\Delta$, this additional term hardly affects the properties of the system. But, when  $\Lambda^{-n/2}\sim\D$, it is not obvious whether the added sites still act as a perturbation in the iterative process (\ref{BCS:H_N+1}) or not,  that is whether the energy scale separation still works or not.
In section \ref{sec:BCS:energyspectrum} we will show that 
the separation of energy scales does work 
also at  resolutions much smaller than the gap.


\subsection{Bogoliubov and particle-hole transformations}
\label{sec:BCS:Bogoliubov}

Satori \emph{et al.}\ \cite{Satori1992} have shown that a computationally more convenient representation of the Hamiltonian can be obtained by performing a Bogoliubov-Valatin transformation [$b_{ln,\s} =   1/\sqrt{2} (\s f_{ln,\s} + f^\dag_{ln,-\s})$]
as well as a particle-hole transformation 
[$\tilde c_{l,2n,\s}=b_{l,2n,\s},\; -\s \tilde c_{l,2n-1,-\s}=b^\dag_{l,2n-1,\s}$]. $n=-1$ represents the dot, thus  $\tilde d_\s=\tilde c_{-1,\s}$.
Applying these transformations to (\ref{BCS:H_N+1}), the Hamiltonian reads
\begeqn
   \tilde{H}_{dot}&=&\frac{ {U}}{2}(1-\tilde n_d+2 \tilde n_{d\up} \tilde n_{d\down}) 
   - ( {\varepsilon_d}+\frac{ {U}}{2})
   (\tilde d^\dag_\up \tilde d^\dag_\down + \tilde d_\down \tilde d_\up)
   \label{BCS:H_dot}
   \\
    \tilde{H}_{hyb}&=&\sqrt{\frac{2 {\Gamma}}{\pi}}\sum_{\sigma}
   \left[\cos{{{\phi}\over {4}}}\;\;(\tilde c^\dag_{e0\s}\tilde d_\s + h.c.)-\sin{{\phi\over 4}}\;\;(\tilde c^\dag_{o0\s}\tilde d_\s + h.c.)\right]
   \label{BCS:H_hyb}
   \\
   \tilde{H}_{lead}
   &= & 
   \frac{1}{2}(1+\Lambda^{-1})  
   \sum_{l=eo}\sum_{\s,n=0}^\infty \Lambda^{-n/2}\,\xi_n \,
   (\tilde c^\dag_{ln\s}\tilde c_{ln+1\s}+\tilde c^\dag_{ln+1\s}\tilde c_{ln\s})
   \label{BCS:H_l}
   \\
    \tilde{H}_{\D}
   &=&
   - \sum_{n=0,\s}^\infty (-1)^n  {\Delta}\;
   (\tilde n_{en\s}-\tilde n_{on\s}).
   \label{BCS:H_BCS}
\eeqn
Operators in the new basis will always be denoted by a tilde, e.g.\ $\tilde{n}_{d\s}=\tilde{d}^\dag_{\s}\tilde{d}_\s$ or $\tilde{n}_{ln\s}=\tilde{c}^\dag_{ln\s}\tilde{c}_{ln\s}$.
The $\tilde{Q}$ nonconserving property of  $H_\D$ has been transferred to  $\tilde H_{dot}$. This has two useful effects: 
(i) For the \emph{symmetric model} not only the $z$-component of the total spin (per iteration), $\tilde S_{zN}=\frac{1}{2}\sum_{l,n=-1}^N (\tilde n_{ln\up}-\tilde n_{ln_\down})$,
 but also the particle number $\tilde Q_N=\sum_{l,n=-1,\s}^N (\tilde n_{ln\s}-1)$ is a conserved quantum number (note that this definition yields $\tilde Q=0$ for the Fermi sea together with a singly occupied dot at $\Gamma=0$).
 Therefore the dimensions  of the matrices to be diagonalized at each iteration (and therefore the numerical effort) is reduced significantly.
(ii) 
Additionally, in the non-interacting symmetric case ($U,\varepsilon_d = 0$) the Hamiltonian takes a very simple quadratic form. We will focus on its exact solution  in the next section.
In section \ref{sec:BCS:energyspectrum} the resulting single-particle picture will  serve as a tool to gain a deeper understanding of  reasons why NRG does work for the SC-AM.

For simplicity, we use $\phi=0$ in the following. Then, the odd channel decouples and the problem reduces to an effective one-lead system. The resulting model is equivalent to that describing an impurity embedded in a bulk superconductor.

\subsection{Single-particle picture}
\label{sec:BCS:U0}

Some properties of the system show up already in the non-interacting case, $U=0$. The Hamiltonian is then of quadratic form and we only have to solve a single-particle problem. The NRG Hamiltonian [(\ref{BCS:H_dot})-(\ref{BCS:H_BCS})] can be diagonalized  up to a large number of iterations exactly - that is without truncating the Hilbert space.
One can use the resulting exact solution as benchmark for the NRG result. 
We obtain very good agreement in the energy spectrum, thus confirming that NRG is capable of accurately treating superconducting leads. 
In section \ref{sec:BCS:energyspectrum} the single-particle picture will also serve as a tool to gain a deeper understanding of  reasons why NRG does work for the SC-AM.

Without lack of generality we  restrict the discussion to the symmetric case,  $\varepsilon_d=0$. Then the NRG Hamiltonian  only contains quadratic terms of the form $a_i^\dag a_{i'}$, with $a_{i}$ some fermionic operator. 
For every iteration $N$ the single-particle Hamiltonian can  be diagonalized by some unitary transformation $T$ to $  H_N=\summ_{j=-1}^N \varepsilon_j   \alpha^\dag_j \alpha_j$. Here $\alpha_j=T^\dag_{ji}a_i$ and the  eigenstates  $|n_j\rangle= \alpha^\dag_j |{\rm vac}\rangle$ satisfy $\alpha^\dag_j\alpha_j |n_j\rangle = n_j |n_j\rangle$. 
The many-body  eigenstates and the energy spectrum follow from the Schr\"odinger equation
\begeq
	H |m\rangle = E_m |m\rangle,
        \,\,\,\,\,
        E_m  = \left(\summ_{\{n_j\}_m} \varepsilon_j n_j\right) - E_0,
        \label{U0:Em}
\eeq
with the many-body eigenstates $|m\rangle=|n_1\dots n_N\rangle$ and eigenenergies $E_m$ which are calculated w.r.t.\ the ground state energy $E_0$.
In the ground state $|0\rangle$ all single-particle levels with energy below the Fermi energy $\varepsilon_F=0$ are occupied, thus $E_0=\summ_{l ( \varepsilon_l<0)} \varepsilon_l$.
Expectation values of local operators are evaluated easily, e.g.\ 
$\langle 0 | a_{-1} a_{-1}^\dag | 0 \rangle
   =
   \sum_l U^\dag_{-1,l} \langle 0 | 
         \alpha_{l} \alpha_{l}^\dag          | 0 \rangle U_{l,-1}
   =
   \sum_{l(\varepsilon_l<0)}  |U_{l,-1}|^2.$

\begin{figure}[h!]
  \centering
  \includegraphics*[width=0.57\columnwidth]{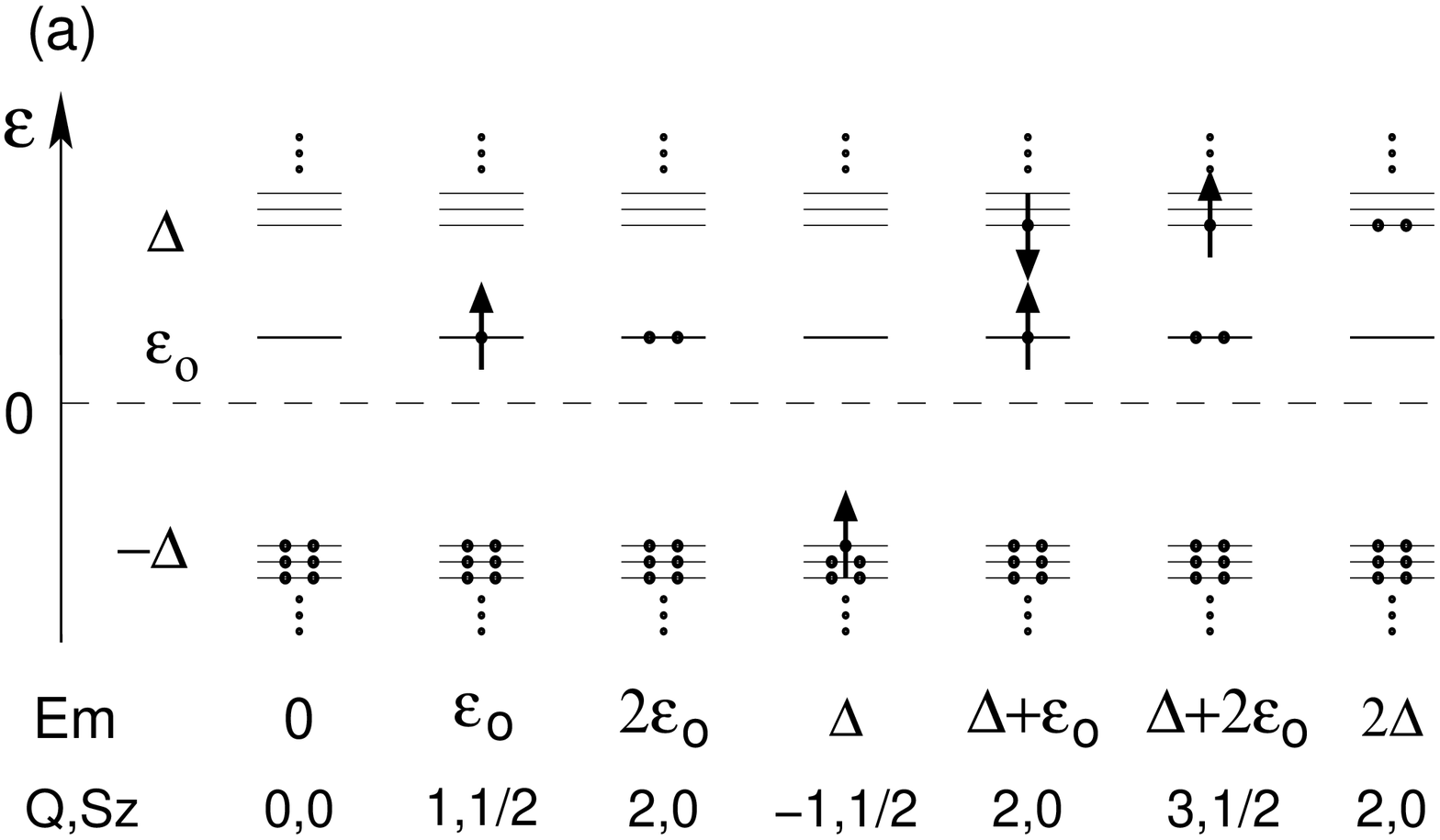}
  \includegraphics*[width=0.41\columnwidth]{figure_1b.eps}
  \caption{ Single-particle and many-body  eigenenergies for the SC-AM ($U=\varepsilon_d=0$). 
    (a) Schematic sketch of a typical single-particle level spectrum of the SC-AM. The continuum is represented by a discrete set of closely-spaced levels. The construction of the lowest lying many-body eigenstates in terms of single-particle states according to (\ref{U0:Em}) is illustrated. The corresponding energies $E_m$ (w.r.t.\ the ground state energy) and quantum numbers of the many-body states are also given. Two dots stand for a doubly occupied level.
    (b) Single-particle energies $\varepsilon_{j}$ ($\varepsilon^{+}_{j}>0$: stars, $\varepsilon^{-}_{j}<0$: crosses), as well as the corresponding many-body eigenenergies $E_m$ (lines) for $\Delta=10^{-4}$ and $\Gamma/\D=0.3$ at a late NRG iteration ($\delta_n\ll \Delta$). The many-body eigenenergies depicted in  (a) are specified on the right hand side.
    The single-particle continuum energies go like $\varepsilon^{\pm}_{j}-\Delta\propto \Lambda^{2j}$, see section \ref{sec:BCS:continuum}. 
    Due to many-particle excitations, this dense profile repeats as substructures in the many-body spectrum at $m\D+r\varepsilon_0$ ($m=1,2,\dots$, $r=0,1,2$).
    
  }
   \label{fig:BCS:U0:excit}
   \end{figure}

The construction of the many-body spectrum from the single-particle energy levels using  (\ref{U0:Em}) is illustrated in figure \ref{fig:BCS:U0:excit} for the  SC-AM.
Figure \ref{fig:BCS:U0:excit}(a) shows a sketch of a typical single-particle spectrum.
The single-particle level spectrum consists of a  continuum  above and below the gap, $|\varepsilon_{l}| > \Delta$
(represented  by a discrete set of closely-spaced levels), as well as one subgap level with energy $0\leq\varepsilon_0<\Delta$, the so-called Andreev level. 
Note that because of the discretized conduction band, we also have a discretized continuum.

The sketch also demonstrates the construction of the lowest lying many-body eigenenergies using (\ref{U0:Em}).
For $\Gamma>0$,  no single-particle level exists at the Fermi energy and the many-body ground state is a singlet ($\tilde S_z=0$, $\tilde Q=0$).
The first excitation  is a degenerate doublet ($E_{1,2}=\varepsilon_0$, $\tilde S_z=\pm 1/2$,  $\tilde Q=1$), corresponding to the bound single-particle level, occupied by either a spin up or down. 
If $ \varepsilon_0<\Delta/2$, an additional subgap  state forms ($E_3=2 \varepsilon_0<\Delta$,  $\tilde S_z=0$, $\tilde Q=2$), corresponding to spin up \emph{and} down occupying the subgap single-particle level.
Otherwise $E_3>\Delta$ is part of the continuum energies.
A concrete example of the many-body as well as the single-particle eigenenergies is shown in  figure \ref{fig:BCS:U0:excit}(b). 
Both the single-particle levels (stars, crosses) as well as the resulting many-body eigenenergies are shown. On the vertical axis the energies of the many-body eigenstates constructed in figure \ref{fig:BCS:U0:excit}(a) are specified.

\section{Energy spectrum}
\label{sec:BCS:energyspectrum}
In this section we analyze the many-body energy spectrum generated during the iterative NRG procedure. As already mentioned in the last section, the spectrum consists of a  continuum above and subgap bound states below the gap $\Delta$. The competition between the Kondo effect and Cooper pair formation is reflected in the ground state properties of the system. 
A detailed analysis of the structure of the continuum with the help of the single-particle picture reveals that, interestingly, energy scale separation is even more efficient at energy scales smaller than the gap (compared to the AM).

\begin{figure}[h!]
  \centering
  \includegraphics*[width=1\columnwidth]{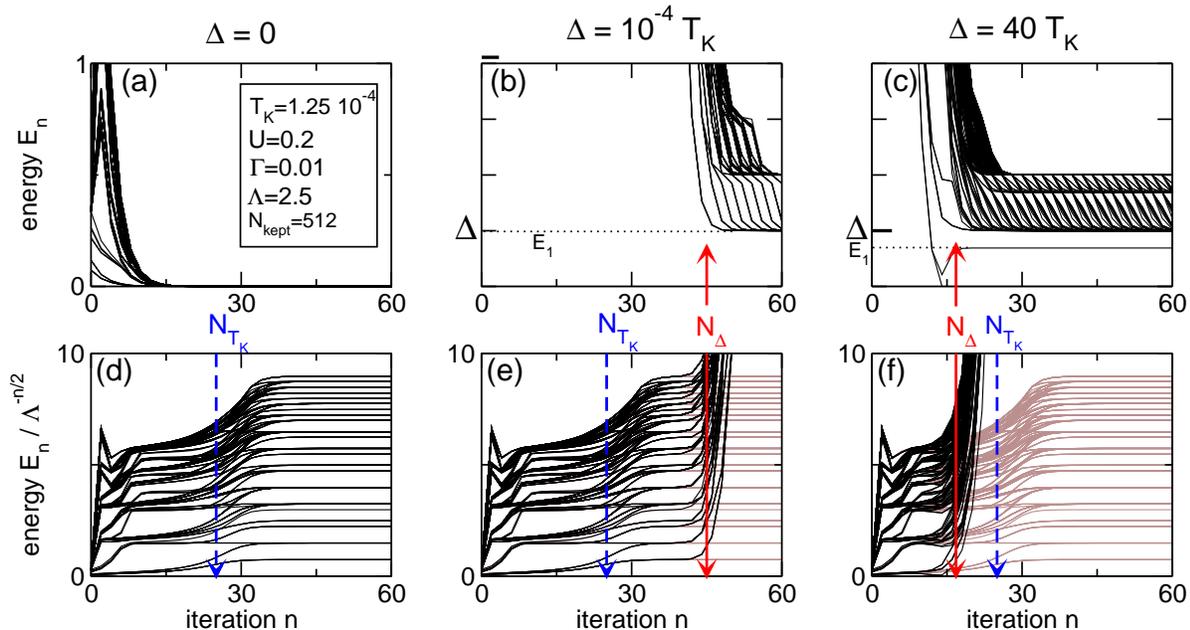}
  \caption{ 
    Energy spectra (a,b,c) and the corresponding energy flow diagrams (d,e,f) for the lowest 280 eigenenergies of the even NRG iterations of a SC-AM. $U=0.2$, $\Gamma=0.01$, $T_K=1.25\cdot 10^{-5}$, $\Lambda=2.5$ and $N_{\rm kept}=512$ in all plots. The iteration numbers $N_{T_{K}}$ or $N_{\D}$, where $\delta_n^{\rm AM}\approx T_{K}$ or $\Delta$, respectively,  are indicated by dashed or solid arrows. 
    (a,d): $\Delta=0$, AM. Since $\delta_n^{\rm AM}\propto\Lambda^{-n/2}$, the eigenenergies (a) fall off exponentially with $n$ and the energy flow diagram (d) converges. At energy scale $T_K$ the localized spin gets screened by the conduction electrons and the Kondo singlet (ground state) forms.
    (b,c,e,f): $\Delta>0$. At energy scale $\Delta$ the exponential decrease crosses over to a saturation towards $\Delta$. The characteristic gap as well as the Andreev bound states form. Consequently, the flow diagram energies grow with $\Delta\Lambda^{n/2}$ due to rescaling. For comparison, the result for the AM is indicated (brown), too.
    For $\Delta\ll T_K$ (b,e) it is energetically favorable to break Cooper pairs and lower the energy by $T_K$ by screening the local spin, thus the ground state is a singlet. For $\Delta\gg T_K$ (c,f) Cooper pair formation dominates the low energy properties and the ground state is  a doublet.
  }
  \label{fig:BCS:flowdiag}
\end{figure}

Figure \ref{fig:BCS:flowdiag} shows the 280 lowest lying many-body eigenstates for the even NRG iterations of a SC-AM in different regimes of $T_{K}/\D$.
We first discuss the case $\Delta=0$ (AM, figure \ref{fig:BCS:flowdiag}(a),(d)), then $\Delta\neq 0$.
For the AM the effective level spacing of the Wilson chain drops exponentially with every added site (see discussion above). The energy resolution of the kept states is enhanced exponentially with increasing iteration $n$, see figure  \ref{fig:BCS:flowdiag}(a). 
Thus, an appropriate way of visualizing the physics at different energy scales is given by the rescaled energy spectrum. In these ``energy flow diagrams'', 
the eigenenergies are plotted in units of  $\Lambda^{-n/2}\propto \delta_n^{\rm AM}$, see figure \ref{fig:BCS:flowdiag}(d). 
Only at energy scales where the system changes its properties,
the flow of the eigenenergies changes.
For the AM we are interested in the lowest of these scales, the Kondo scale
 $T_K=\sqrt{\frac{U\Gamma}{2}}\exp\left[\frac{\pi\epsd}{2\Gamma U}(U+\epsd)\right]$, indicated by dashed arrows in the figure.
For details of the various fixed points of the AM see e.g.\ \cite{Krishnamurthy1980}.

In contrast to the exponential decaying couplings of the Wilson chain (\ref{H_l_NRG}),  the  on-site contribution $\Delta$ of the pairing term  (\ref{H_BCS_NRG}) is constant in magnitude for each site. 
Consequently, for $\delta_n<\D$, the BCS contribution is a relevant perturbation and determines the physics of the system. 
Typical energy spectra (or energy flow diagrams, respectively) for finite $\D$ are shown in figure \ref{fig:BCS:flowdiag}(b,c) (or \ref{fig:BCS:flowdiag}(e,f)). 
At energy scale $\Delta$, the exponential reduction of the eigenenergies crosses over to a saturation towards $\Delta$. The characteristic gap as well as the subgap Andreev bound states form.

The structure of the continuum energies near the gap will be discussed in section \ref{sec:BCS:continuum}. We show there that even though the on-site terms of $H_\Delta$ do not fall off exponentially like the couplings of the Wilson chain, the energy scale separation (the heart of NRG) still works. 


\subsection{Competition between Kondo effect and Cooper pair formation}
\label{sec:BCS:competition}

The system  possesses two  energy scales  determining the characteristics of the system, $T_K$ and $\D$. 
The resulting competition between Kondo effect and Cooper pair formation is reflected in the ground state properties of the system: 
For $T_K\ll\Delta$, $H_{\D}$ is the dominant term of the Hamiltonian. The system lowers its energy by $\Delta$ by the formation of Cooper pairs leading to effective spin zero bosons (singlets) in the reservoir. The lead density of states gets depleted, a gap from $-\Delta$ to $\Delta$  forms.
 Therefore  no electrons are available to screen the localized spin. Consequently, the ground state is a spin doublet with $S_z=\pm1/2$.
In case when the Kondo effect is dominant $(T_K\gg\Delta)$, it is energetically favorable to break Cooper pairs so that
the localized spin on the quantum dot gets screened by the non-paired electrons near the Fermi energy. The energy is lowered by $T_K$ and as ground state the typical Kondo singlet forms $(S_z=0)$.
The influences of the different scales is
also apparent in  the NRG energy flow diagrams, see figure \ref{fig:BCS:flowdiag}(e),(f). For example, for $T_K\gg\Delta$, the effect of $H_\Delta$ sets in at iterations after 
the Kondo fixed point is reached (i.e.\ at lower energies).
A phase diagram for the singlet and doublet ground state including the spectral weight of the bound states was recently derived by \cite{Bauer2007} (also using NRG) for the whole regime of $\Delta$, $\Gamma$.

\subsection{Analysis of the continuum}
\label{sec:BCS:continuum}

The key feature of NRG is the energy scale separation: The couplings between successive sites of the Wilson chain describing a normal lead fall off exponentially, therefore each 
added site can be treated as a perturbation of the previous system. 
However, when generalizing the AM to superconducting leads, a \emph{constant} on-site energy $\Delta$ is added at each site (see (\ref{BCS:H_N+1})). 
In order to understand why NRG works even at resolutions well below $\Delta$, we now take a closer look at the structure of the continuum produced during the iterative NRG procedure.

\begin{figure}[h!]
  \centering
  \includegraphics*[width=0.5\columnwidth]{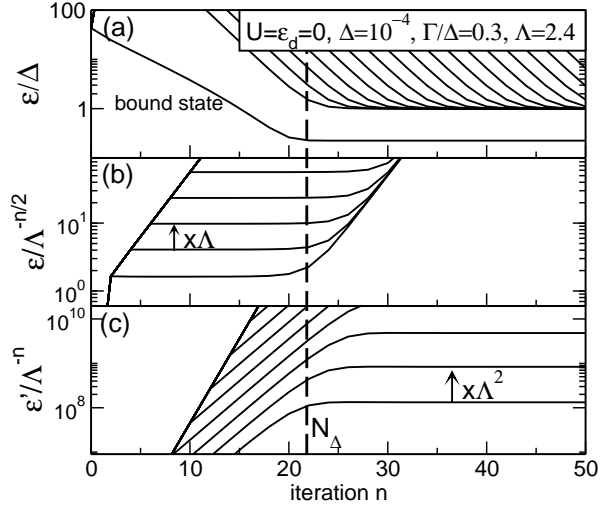}
  \caption{ Positive single-particle energies plotted for three different scalings of the vertical axis for even iterations. The vertical dashed  line indicates the iteration $N_\D$ for which $\delta_n^{\rm AM}\approx\D$. (a) Energy versus iteration number. The single-particle levels flow towards the gap and form a continuum. The bound state can be also seen. 
    (b) The eigenenergies $\varepsilon$ representing the continuum in units of $\Lambda^{-n/2}$, corresponding to the customary scaling for energy flow diagrams of the (normal) AM.
    For $n<N_\D$, the single-particle energies run horizontally, since they obey  $\varepsilon_{jn}\propto \Lambda^{j-n/2}$, see text and \cite{Krishnamurthy1980}.
    (c) $\varepsilon '$ in units of $\Lambda^{-n}$. The single-particle energies run horizontally for $n>N_\D$, since here they scale as   $\varepsilon_{jn}'\propto \Lambda^{2j-n}$. The line sloping upwards at the left of figures  (b) and (c) is due to the fact that  at every iteration a degeneracy is split and hence an extra eigenenergy is generated.
  }
  \label{fig:SP_BCS_U0_D0.0001_G0.3_L2.4}
\end{figure}

We therefore analyze the (positive) continuum of the single-particle problem.
Figure \ref{fig:SP_BCS_U0_D0.0001_G0.3_L2.4} shows an example of a single-particle spectrum for three different scalings of the  vertical axis: In (a) the  (unscaled) spectrum is plotted versus the iteration number. The NRG eigenenergies decrease with iteration number $n$ and tend towards $\Delta$.
However,
the decrease of the continuum eigenenergies
depends on whether $n$ is smaller or larger than $N_\Delta$, the iteration number for which $\delta_n^{\rm AM}\approx \Delta$.
We find the following asymptotic behaviour:
\begeq
\varepsilon_{jn}:\,\,\,\left\{\matrix{
   \varepsilon_{jn}&\propto& \Lambda^{j-n/2}&{\mbox{    for    }}& \,\,n<N_\D,
   \cr
   \varepsilon_{jn}'=\varepsilon_{jn}-\Delta&\propto& \Lambda^{2j-n}\,\,\,\,\,&
   {\mbox{    for    }}& \,\,n>N_\D.
   }
   \right.
   \label{BCS:epsilon_k}
\eeq
Primed energies will henceforth always be understood to be measured relative to $\Delta$.
Both relations of (\ref{BCS:epsilon_k})  are illustrated in figures \ref{fig:SP_BCS_U0_D0.0001_G0.3_L2.4}(b) and \ref{fig:SP_BCS_U0_D0.0001_G0.3_L2.4}(c),  by plotting the eigenenergies $\varepsilon_{jn}$ and $\varepsilon_{jn}'$ in units of $\Lambda^{-n/2}$ and 
$\Lambda^{-n}$, respectively.

\begin{figure}[h!]
    \centering
    \includegraphics*[width=0.4\columnwidth]{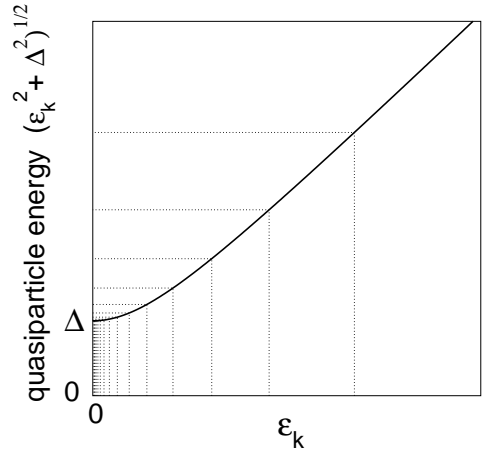}
    \caption{  Logarithmic discretization of $\varepsilon_k$ leads to high resolution at the band edge of the BCS quasiparticle eigenenergies.
         }
    \label{fig:BCS:discretization}
    \end{figure}

These results can already be understood by further reducing the problem to only the superconducting reservoir ($H_l+H_\Delta$):
For fixed iteration $n$, the $j$-dependence of  (\ref{BCS:epsilon_k}) 
reflects  the standard logarithmic discretization of the continuous conduction band, according to which the single-particle energies of the Wilson chain grow  in powers of $\Lambda$, i.e.\  $\varepsilon_k \propto\Lambda^j$ \cite{Krishnamurthy1980}.  Inserting this into the single-particle dispersion relation of BCS quasiparticles, the effective discretization of the isolated superconducting reservoir is obtained, as sketched in figure \ref{fig:BCS:discretization}. The limits yield
\begeq
   \xi_k=\sqrt{\varepsilon_k^2+\D^2}
   :\,\,\,
   \left\{
   \matrix{
     \xi_k\approx&
     \varepsilon_k &\rightarrow& \Lambda^j & \mbox{for}&  \varepsilon_k\gg\D
     & ( n<N_\D),
     \cr
     \xi_k'\approx&
     \frac{\varepsilon_k^2}{2\D} &\rightarrow&\Lambda^{2j}& \mbox{for}&  \varepsilon_k\ll\D
     &(n>N_\D ).
   }
   \right.
   \label{BCS:continuum_j}
 \eeq
The $n$-dependence of   (\ref{BCS:epsilon_k}) follows heuristically from considering the coupling of two neighbouring sites $n-1$ and $n$, with hopping matrix element $t_n\sim \delta_{n}^{\rm AM}$ and on-site energy $\pm \D$  (see  (\ref{BCS:H_l}) and (\ref{BCS:H_BCS}), respectively).
Since $t_n\sim\Lambda^{{-n/2}}$, the eigenstates $\lambda_{\pm}$ of this two-state problem show the asymptotic behaviour
\begeq
   \lambda_{n\pm}=\pm\sqrt{\D^2+t_n^2}:\,\,\,
   \left\{
   \matrix{
     \lambda_{n+}\approx &t_n &\propto &\Lambda^{-n/2} &\,\, \mbox{for}&  n<N_\D,
     \cr
     \lambda_{n+}'\approx &\frac{t_n^2}{2\D} &\propto& \Lambda^{-n} & \,\,\mbox{for}&  n>N_\D.
   }
   \right.
\eeq

Note that the $\Lambda^{-n/2}$ versus $\Lambda^{-n}$ scaling of the two limits of  (\ref{BCS:epsilon_k}) implies that the energy scale separation is actually \emph{more} efficient in the second limit, when the gap dominates the spectral features.
\emph{This establishes one of the central results of the present paper:}
The energy scale separation, being the heart of the  NRG approach, is not impaired but rather enhanced (w.r.t.\ $\Delta$) by the presence of the energy gap in the superconducting leads.
Increasing the chain length leads to an exponential enhancement of the resolution at the continuum edge at $\Delta$.

The many-body spectrum is constructed according to equation (\ref{U0:Em}). For $n<N_\D$, as in the AM, the mean level spacing (at fixed iteration) does not depend on energy, as can be seen in figure \ref{fig:BCS:flowdiag}(d-f). 
For $n>N_\D$ the gap forms, and every many-body state with energy $E_m<\D +
\varepsilon_0$ can only stem from  adding one electron (or hole) to the Fermi sea. $\varepsilon_0$ is the energy of the single-particle subgap level.
Therefore in this regime $E_{j\s n}= \varepsilon_{jn}$. 
Due to many-particle excitations, this dense profile repeats as substructures in the many-body spectrum at $m\D+r\varepsilon_0$ ($m=1,2,\dots$; $r=0,1,2$), as can be found in figure \ref{fig:BCS:U0:excit}(b), which was calculated  for some late iteration with $n\gg N_\Delta$.
%

%

\section{Spectral function}
\label{sec:BCS:A}

In this section we study basic properties of the  spectral function that are special for the SC-AM. We therefore begin by  analyzing the analytic solution of the non-interacting case and find not only a continuum for energies $|\w|>\Delta$ together with subgap resonances (as expected from the energy spectrum), but  also  a sharp peak at the gap edge for $\Delta<\Gamma$.
The agreement between NRG results and the analytic solution is excellent, especially when resolving that sharp feature, again confirming that NRG is valid also at resolutions well below the gap. Subsequently, we consider spectral functions at finite $U$, which show some of the same feature as found for the non-interacting case.

Knowledge of the energy spectrum suffices to calculate thermodynamic quantities, such as the impurity specific heat. We focus here instead on the more complex calculation of the local spectral function $A_\s(\omega)$.
As this is a dynamic quantity, all energy scales have to be taken into account even at temperature zero.
The spectral function is defined as 
\begeq
\mathcal{A}_\s(\omega) 
   =
   -\frac{1}{\pi} {\rm Im}\, G^R_\s(\omega),
   \label{A_ImG}
\eeq
where $G^R_\s(\omega)$ is the Fourier-transformed of the retarded Green's function 
$G^R_\s(t)\equiv -i\theta (t) \langle [ d_\s (t),d_\s^\dag (0) ]_+ \rangle$.
Motivated by the structure of the eigenspectrum we distinguish two contributions to the spectral function,
 \begeq
   \mathcal{A}_\s(\omega) =\sum_{|E_m|<\Delta} w_m \delta(\omega-E_m) 
   + A_\s(\omega)\,\theta(|\w|-\D).
   \label{BCS:A_dr}
\eeq
$w_m$ denotes the weight of excitations to the Andreev bound states, which contribute as $\delta$-peaks within the gap between $-\D$ and $\D$. $A_{\s}(\w)$ represents the continuum contribution, with $|\w|>\D$.
In the following we present results for the continuum part of the spectral function. We focus on the behaviour of the continuum contribution of the spectral function close to the gap.

\subsection{NRG} 

NRG calculations of the spectral function are based on the Lehmann representation:
\begeqn
   \mathcal{A}_\s(\omega) 
   &=&
   \frac{1}{Z}\sum_{mm'}
   (e^{-\frac{E_m}{k_BT}}+e^{-\frac{E_{m'}}{k_BT}})
   \,\,|\left<m|         
     d^\dag_{\s}
     |m'\right>|^2 
   \,\, 
   \delta(\omega-(E_{m}-E_{m'})).
   \label{BCS:A}
\eeqn
Here  $Z=\sum_m e^{-E_m/k_BT}$ is the partition function at temperature $T$, and $|m\rangle$ denotes an exact eigenstate of the Hamiltonian with eigenenergy $E_m$.
The matrix elements of these operators as well as the eigenenergies can be calculated with NRG at all energy scales.
The $\delta$-peaks of the continuum contribution are broadened as described by \cite{Bulla2001,Weichselbaum2007}.
As expected from the findings about the spectra, broadening w.r.t.\ $\Delta$ (i.e.\ using $\omega^\prime=|\omega|-\Delta$ in \cite{Weichselbaum2007}) leads to good results, see below.
We use the full density matrix NRG \cite{Weichselbaum2007,Peters2006} at effective temperature zero, i.e.\ at temperature  much smaller than any other energy scale of the problem. Only matrix elements connecting the ground state(s) with excited states then contribute.

As NRG parameters we choose $\Lambda=1.8$ and
use $z$-averaging \cite{Oliveira1994} (where for fixed $\Lambda$ data is averaged for different discretizations) with an interval spacing of $\delta z=0.05$ to impoove the results. We keep ${ N_{\rm kept}} = 1024$  at the first $6$ to $10$ iterations, and use  $N_{\rm kept}=512$ for the rest of the iterations.
We test NRG against the analytic solution at $U=0$ and find excellent agreement, implying that the  broadening procedure as well as the choice of $ N_{\rm kept}$ are adequate for the present problem.
%

%

\subsection{Spectral function  for  $U=0$}
\label{sec:BCS:A_U0}


\begin{figure}[h!]
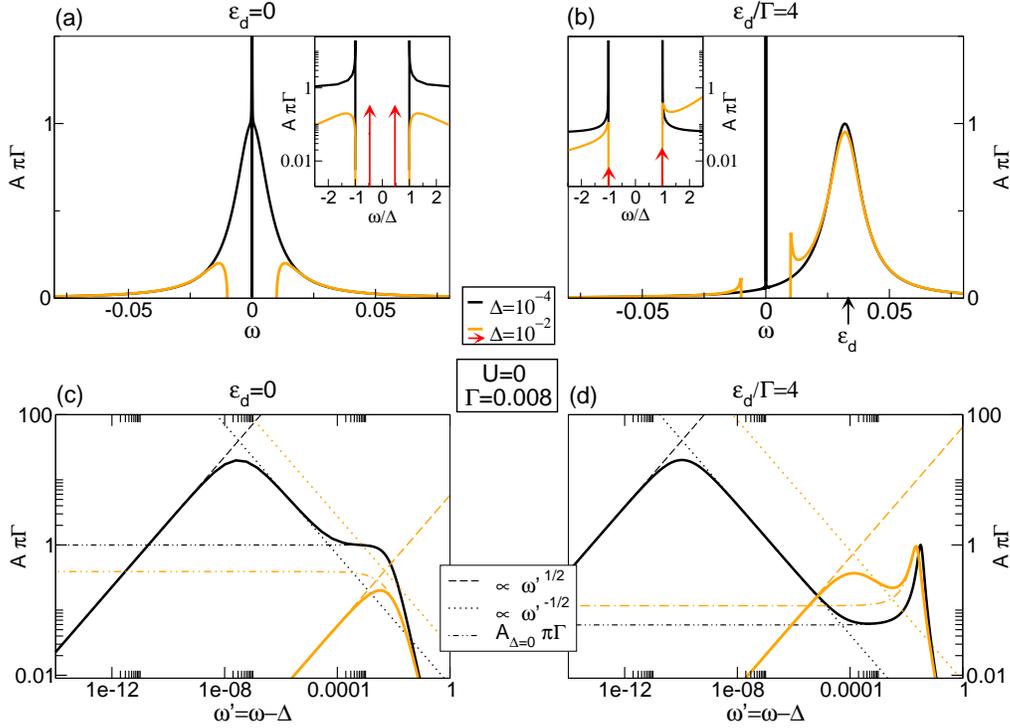

  \centering
  \includegraphics*[width=0.85\columnwidth]{figure_5ab.eps}
  \includegraphics*[width=0.85\columnwidth]{figure_5cd.eps}
  \caption{ 
    Continuum contribution $A(\w)$ to the spectral function for $U=0$, $\Gamma=0.008$ and $\Delta=10^{-4}$ as well as $\Delta=10^{-2}$ obtained from  equation (\ref{BCS:A0}). (a,c) show a symmetric and (b,d) an asymmetric SC-AM. 
In (a,b), a linear scale is used revealing sharp peaks near the gap if $\omega_c ' \ll \Delta$. The insets, which zoom in the region of the gap edge, show the full height of the near-gap peaks. They also indicate, using position and length of arrows, the energy and weight of the subgap contribution for the case $\D=10^{-2}$ (calculated with NRG).  In (c,d), for the same data a log-log scale is used to elucidate the asymptotic behaviour of  (\ref{BCS:A:tangent}) (dashed and dotted lines) and (\ref{BCS:A:tangent_2}) (dashed-dotted lines).
  }
  \label{fig:BCS:A_U0_Asympt}
\end{figure}        


\begin{figure}[h!]
  \centering
  \includegraphics*[width=0.85\columnwidth]{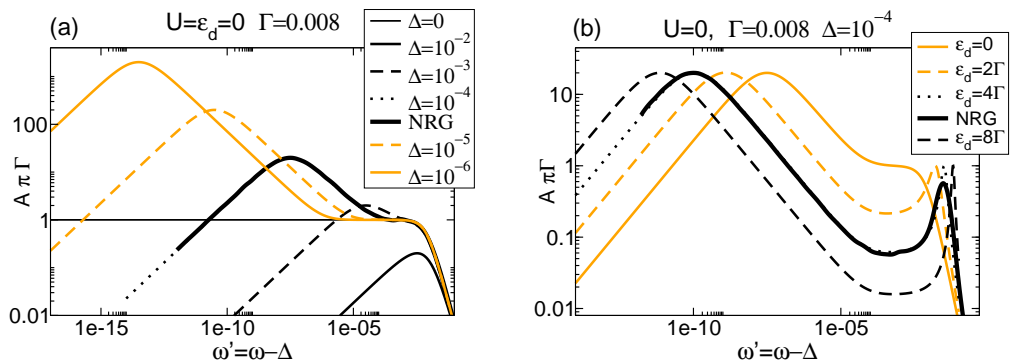}
  \caption{ 
    Continuum contribution $A(\w')$ to the spectral function for $U=0$, $\Gamma=0.008$, as obtained from  (\ref{BCS:A0}), for (a) various values of $\Delta$ and $\epsd=0$ (symmetric model) and (b) $\Delta=10^{-4}$ and various values of $\epsd$ (asymmetric model).  For $\w'_{c}\ll\D$   a sharp peak forms at the gap edge.
    The increase (decrease) of  $A(\omega^\prime)$  goes as $\omega'^{\frac{1}{2}} \ \ ( \omega'^{-\frac{1}{2}})$.
    NRG results, shown for $\Delta=10^{-4}$ only (thick black lines), are in excellent agreement with the exact analytical results for $\omega\prime\ll\Delta$.
    Note in (b) that the height of the near-gap peak at the gap edge does not depend on $\epsd$. 
  }
  \label{fig:BCS:A_U0_G0.008_analyt}
\end{figure}

For analyzing the basic properties of $A(\w)$, we first review the non-interacting problem.
There, the local (retarded) Green's function is known exactly \cite{Bauer2007}.
The continuum contribution to $G^{0}$ then reads
\begeq
   G^0_\Delta (\omega)= 
   \frac{1}{D(\omega)}\ 
   \{ (\omega+\epsd) +i\; \Gamma \rho_\Delta \},
   \label{BCS:G0D}
\eeq
with  
$   D(\omega) = 
   (\omega^2 - \Gamma^2 - \epsd^2) + i\; 2\Gamma \omega \rho_\Delta
$,
    and 
 $\rho_\D$  the density of states of a bulk superconductor. At temperature $T=0$, the latter  has the limits
 \begeq
   \rho_\Delta(\omega)
   =
   \frac{\theta(|\w|-\D)\,|\omega|}{\sqrt{\w^2-\D^2}}
   \approx
   \left\{
   \matrix{ 
     \sqrt{\frac{\Delta}{2{|\w'|}}}\,\,\theta(\w'), & {\rm for } & \w'\ll \Delta,
     \cr
     1, & {\rm for } &\D \ll |\w|.}
   \right.
         \label{BCS:rhoD}
\eeq
For $\D=0$, equation (\ref{BCS:G0D}) simplifies to the well known formula for the AM, 
$G^0_{\D=0}(\w)= (\w-\epsd-i\Gamma)^{-1}$. 
For the spectral function we 
obtain from  equations (\ref{BCS:G0D}) and (\ref{A_ImG}), 
\begeq
   A(\omega)
   =
      \frac{(\w+\epsd)^2+\G^2}{(\w^2-\epsd^2-\G^2)^2 + (2\G^2\w\rho_\D)^2}
      \,\,\Gamma \,\rho_\D/\pi.
   \label{BCS:A0}
\eeq
This function  is shown for various parameter combinations in figure \ref{fig:BCS:A_U0_Asympt}.
The common features are (i) the atomic resonance of width (half width half maximum) $\sim \Gamma$ centered at $\epsd$, reflecting the level broadening due to the level-lead coupling and (ii) a gap from $-\Delta$ to $\Delta$, with (iii) bound states at some energy $\pm\omega_B$ inside the gap. The energy and weight of the subgap contribution is indicated for $\Delta=10^{-2}$ by position and length of the red arrows in the insets of figure \ref{fig:BCS:A_U0_Asympt}(a,b). 
Here they are calculated with NRG, but can be also obtained analytically, see equation (7) of \cite{Bauer2007}.
Note that for finite $\Gamma$ bound states exist also for $\epsd\gg\Delta$, see figure \ref{fig:BCS:A_U0_Asympt}(b). They asymptotically approach the gap edge for $|\epsd| \rightarrow \infty$, in accordance with equation (7) of \cite{Bauer2007}. 

Additionally, the continuum part of the spectral function may feature near-gap sharp peaks, point of interest in the following discussion.
The behaviour near the gap edge can be approximated (using equation (\ref{BCS:rhoD}) and writing $s=\rm sign(\w)$) by
\begeqn
   A(\omega)
   &\approx&
   \frac{(s\D+\epsd)^2+\G^2}{(\D^2-\G^2-\epsd^2)^2 + 4\G^2\D^2 + 2\G^2\D^3/\w'}\,\Gamma \rho_\D/\pi
   \label{BCS:A:approx}
   \\
   &\approx&
   \left\{
     \matrix{
       \frac{(s\D+\epsd)^2 +\G^2}{2\G^2\D^3}
       \; \G \rho_\D \w'/\pi
       &
       \propto \sqrt{\w'},
       & 
       {\rm for }
       &
       \w'\ll \w'_c,
       \cr
        \frac{(s\D+\epsd)^2+\G^2}{(\D^2-\G^2-\epsd^2)^2 + 4\G^2\D^2}
       \; \Gamma\rho_\D /\pi
       &
       \propto \frac{1}{\sqrt{\w'}},
       &
       {\rm for }
       &
       \w'_c\ll \w'\ll\Delta.
       \cr
       }
     \right.
     \label{BCS:A:tangent}
\eeqn
The limits given in (\ref{BCS:A:tangent}) are indicated by the thin lines in figure \ref{fig:BCS:A_U0_Asympt}(c,d): 
$A(\w)$ increases as $\sqrt{\w '}$ when $\w '$ is increased from $0$ (dashed lines), decreasing again for $\w'>\w'_c=\frac{2\G^2\D^3}{(\D^2-\G^2-\epsd^2)^2 + 4\G^2\D^2}$ (which is the zero of  derivative of  equation (\ref{BCS:A:approx})). If $\w_{c}'\ll\Delta,\Gamma$, this leads to a very sharp near-gap peak which decreases as $\rho_\D$ (dotted lines). 
Then the near-gap spectral function is greatly enhanced compared to the AM (where the symmetric case yields $A(0)\pi\Gamma =1$).

The solution of the AM (dash-dotted lines) describes the high-energy limit of the SC-AM:
\begeq
   A(\w)
   \approx
       \frac{\G/\pi}{(\w-\epsd)^2+\G^2}
       \,
       = \,
       -\frac{1}{\pi}{\rm Im}\,G^0_{0}
       \,\,\,\,\,\,\,\,\,{\rm for }\,\,\,\,\,
       \D<|\w|.
       \label{BCS:A:tangent_2}
\eeq


The emergence of the near-gap  peak is depicted in figure \ref{fig:BCS:A_U0_G0.008_analyt}(a) for the symmetric model, where the gap $\D$ is varied over four orders of magnitude, starting from $\D=0$.  For $\Delta=10^{-4}$, we also show numerical NRG results (fat solid line). 
Their agreement with the analytical results is excellent.
The height of the near-gap peak does not depend on $\epsd$, see figure \ref{fig:BCS:A_U0_G0.008_analyt}(b), where $\epsd$ is increased up to $8\Gamma$ for fixed $\Delta=10^{-4}$.

\subsection{Spectral function for finite $U$}
\begin{figure}[h!]
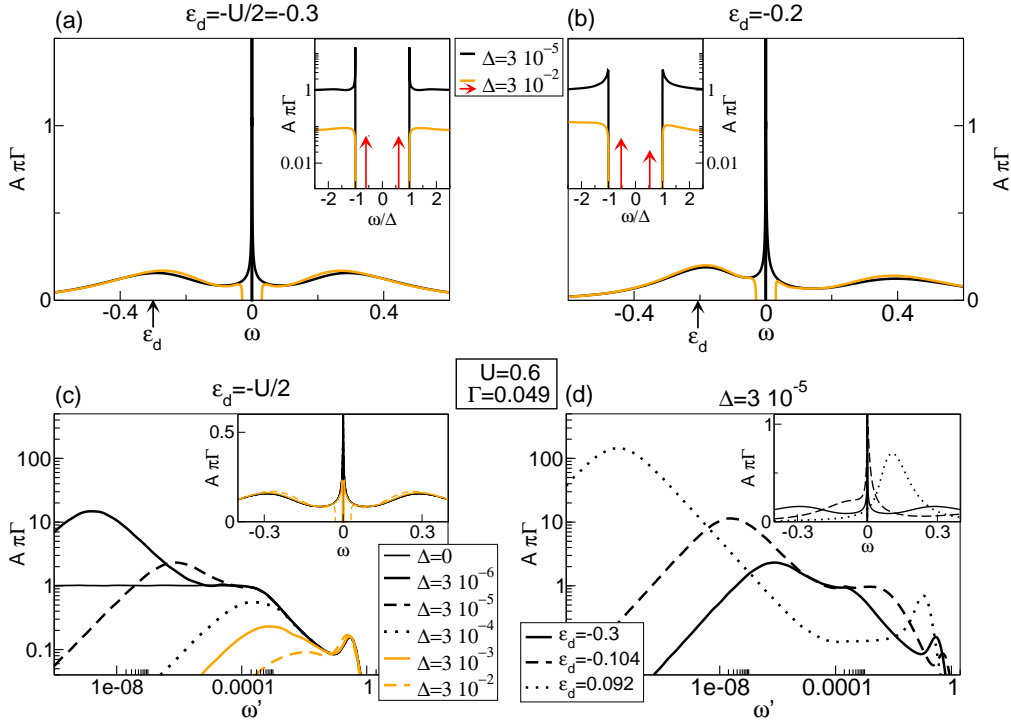

  \centering
  \includegraphics*[width=0.85\columnwidth]{figure_7ab.eps}
  \includegraphics*[width=0.85\columnwidth]{figure_7cd.eps}
  \caption{ Continuum contribution $A(\w)$ to the spectral function for  $U=0.6$, $\Gamma=0.049$, calculated using NRG. 
    (a,c) show a symmetric and (b,d) an asymmetric SC-AM. 
    In (a,b), a linear scale is used revealing the sharp near-gap peaks appearing for $\Delta\ll T_{K}$. The position $\epsd$ of the local level is indicated by an arrow.  
The insets zoom into the gap edge and show the sharp peaks as well as the subgap resonances (for $\D=3\cdot 10^{-2}$, indicated by arrows).
        In (c,d), $A(\w')$ is plotted on a log-log scale for various parameters. The asymptotic behaviour of the near-gap peaks is in agreement with that given in (\ref{BCS:A:tangent}) and (\ref{BCS:A:tangent_2}). In particular, for $\w'\rightarrow 0$, the continuum edge decreases as $\sqrt{\w'}$. 
    (c) $\epsd=-U/2$, $T_{K}=10^{-3}$, various $\Delta$. The singlet-doublet transition occurs between $\Delta/T_K=0.3$ and $3$ (orange: ground state is doublet, black: singlet).
    (d) $\D=3\cdot 10^{-5}$, various $\epsd$, $\D\ll T_{K}$ always.
  }
  \label{fig:BCS:A_U0.6_G0.049}
\end{figure}        

Figure \ref{fig:BCS:A_U0.6_G0.049} shows spectral functions of the SC-AM for finite $U$ and $-U <\epsd < 0$. The two atomic resonances of width $\Gamma$ are now separated by the Coulomb repulsion, thereby they are centered near $\epsd$ and $\epsd+{U}$.
The Coulomb repulsion also drives the Kondo effect, yielding a sharp resonance of width $T_{K}$ pinned at the Fermi energy. 
This resonance is cut by a gap reaching from $-\Delta$ to $\Delta$, reflecting the influence of the superconducting lead.
Depending on the ratio $T_{K}/\Delta$, the Kondo resonance can be cut completely ($T_{K}/\Delta\ll 1$) or emerge clearly ($T_{K}/\Delta\gg 1$).

Additionally, a similar near-gap feature as found for  the non-interacting case may emerge. For $T_{K}/\Delta\gg 1$, a sharp resonance forms at the gap edge, highly exceeding the Kondo resonance, see figure \ref{fig:BCS:A_U0.6_G0.049} for the symmetric case (where the height of the Kondo resonance is given by $1/\pi\Gamma$) as well as the antisymmetric case. The asymptotic behaviour of the near-gap peak is in agreement with that given in equations (\ref{BCS:A:tangent}) and (\ref{BCS:A:tangent_2}). In particular, for $\w'\rightarrow 0$, the continuum edge decreases as $\sqrt{\w'}$.

\section{Conclusion}

The NRG is a well established method for a variety
of quantum impurity models. It is usually applicable in
the whole parameter range and allows to calculate
physical quantities for a wide range of temperatures
and frequencies. In this paper we showed that in the
presence of a superconducting reservoir, NRG provides information for resolutions far below the energy scale of the gap $\Delta$. Moreover,
Wilsonian energy scale separation, being the heart of the success of the NRG approach, is not impaired but rather enhanced by the presence of the energy gap of the superconducting leads. 
This allows  sharp features  of spectral functions at the continuum gap edge to be resolved.
Our calculations of the impurity spectral function cover the  whole region from $\Delta\gg T_K$  to  $\Delta\ll T_K$. In the latter case, we find a sharp peak at the continuum gap edge, vastly exceeding the Kondo resonance contribution.
We expect this to result in an enhanced linear  conductance, as recently reported for experiments with carbon nanotube quantum dots coupled to superconducting leads \cite{Buitelaar2002}.


The ability of the NRG to resolve spectral functions
at energy resolutions well below the gap should be useful for
other problems as well. As discussed in detail in \cite{Fritz2005}, the problem of an impurity in a superconducting
host can be mapped (under certain conditions) to a model
in which the superconductor couples to a normal metal,
with a modified density of states. For the problem
studied in this paper we have performed such a mapping;
the resulting Hamiltonian is given in Equation (\ref{H_BCS_NRG}). In this
case, the oscillating on-site energies, $(-1)^n \Delta$,
generate a hard gap of width $2\Delta$.

This connection allows us, in principle, to calculate the
dynamic quantities for impurity models with arbitrary
gapped bath spectral function (but for this one still
has to develop an algorithm which produces directly
the chain parameters from the hybridization function).
Such calculations might also help to improve the resolution
of the NRG in DMFT calculations for the Hubbard model
where the standard implementation of the NRG does not
describe the shape of the Hubbard bands properly (for
dynamic DMRG calculations for this problem, see 
\cite{Karski2005}).

\ack
We acknowledge helpful discussions with G.\ Zar\'{a}nd and A.\ T\'oth.
This research was supported by the DFG through SFB 484 (RB), SFB-TR12 and De-730/3-2 (AW) and SFB 631 (TH,AW).
Financial support of the German Excellence Initiative via the ``Nanosystems Initiative Munich (NIM)'' is greatefully acknowledged.


\end{document}